# Stable one-dimensional periodic waves in Kerr-type saturable and quadratic nonlinear media

Yaroslav V. Kartashov,[1,2] Alexey A. Egorov,[2] Victor A. Vysloukh,[3] and Lluis Torner[1]

[1]*ICFO-Institut de Ciències Fotòniques, and Department of Signal Theory and Communications, Universitat Politecnica de Catalunya, E-08034 Barcelona, Spain*

[2]*Physics Department, M. V. Lomonosov Moscow State University, 119899, Moscow, Russia*

[3]*Departamento de Fisica y Matematicas, Universidad de las Americas - Puebla, Sta. Catarina Martir, 72820, Puebla, Cholula, Mexico*

We review the latest progress and properties of the families of bright and dark one-dimensional periodic waves propagating in saturable Kerr-type and quadratic nonlinear media. We show how saturation of the nonlinear response results in appearance of stability (instability) bands in focusing (defocusing) medium, which is in sharp contrast with the properties of periodic waves in Kerr media. One of the key results discovered is the stabilization of multicolor periodic waves in quadratic media. In particular, dark-type waves are shown to be metastable, while bright-type waves are completely stable in a broad range of energy flows and material parameters. This yields the first known example of completely stable periodic wave patterns propagating in conservative uniform media supporting bright solitons. Such results open the way to the experimental observation of the corresponding self-sustained periodic wave patterns.

## 1. Introduction

Periodic wave structures in general play a central role in the field of nonlinear waves of different nature [1-7]. They are at the core of the development of modulational instabilities and optical turbulence in continuous nonlinear media, the eigenmodes of such quasi-discrete or discrete systems as waveguide arrays or mechanical, molecular and electrical chains. They are closely related to Bloch waves in solid-state physics, matrices of ultracold atoms or trapped Bose-Einstein condensates, and light-induced reconfigurable photonic lattices. Also self-sustained periodic wave structures were studied in the context of Langmuir plasma waves [8,9], deep-water gravity waves [5,10], pulse trains in optical fibers [1-4,11-17], reconfigurable beam arrays in photorefractive crystals [7,18,19], matter waves in trapped Bose-Einstein condensates [20-24], or synchronously pumped optical parametric oscillators [25,26], to mention a few. By and large, the concept of periodic nonlinear waves allows to bridge the gap between localized solitons and linear harmonic or continuous waves [1,2].

However, periodic wave structures propagating in single-pass conservative media tend to suffer strong instabilities and thus quickly self-destroy. To date stable periodic patterns were known only in physical settings modeled by self-defocusing nonlinearities [3,8-11,27-29], in systems trapped by external potentials, like Bose-Einstein condensates in periodical traps [20-22], or in dissipative confined systems, like optical cavities. To the best of our knowledge, all other known examples of periodic waves in conservative uniform medium supporting bright solitons were found to be dynamically unstable. Actually, the stabilization of periodic waves goes far beyond the properties of each individual structural element, because the stability of the periodic wave is a property of the structure as a whole. Therefore, elucidation of stable periodic waves requires the identification of suitable physical settings where such stabilization might occur, as well as the development of new mathematical techniques to study the stability of periodic nonlinear waves.

Here we review the properties of recently discovered completely linearly stable and metastable (i.e., weakly unstable, but surviving in the numerical simulations for distances exceeding the feasible crystal lengths in the presence of input noise superimposed on the exact stationary solutions) periodic waves in Kerr-type saturable and quadratic nonlinear media, supporting bright and dark localized solitons. We reveal that in both cases, stabilization of bright periodic waves occurs above certain threshold power level. We also show that, in contrast to the results known for pure Kerr nonlinear media, dark periodic waves can be destabilized by saturation of nonlinear response, while dark quadratic waves turn out to be metastable in the broad range of material parameters. Linear stability analysis of periodic waves was revealed by a specific new mathematical formalism, based on a Floquet approach, and was confirmed by direct numerical simulations of propagation of the periodic waves perturbed with white input noise.

## 2. Periodic waves in saturable media

Propagation of optical radiation in (1+1) dimensions in saturable Kerr-type medium is described by the nonlinear Schrodinger equation for the slowly varying field amplitude $q(\eta,\xi)$:

$$i\frac{\partial q}{\partial \xi} = -\frac{1}{2}\frac{\partial^2 q}{\partial \eta^2} + \frac{\sigma q|q|^2}{1+S|q|^2}. \tag{1}$$

Here transverse $\eta$ and longitudinal $\xi$ coordinates are scaled in terms of the characteristic pulse (beam) width and dispersion (diffraction) length, respectively; $S$ is the saturation parameter; $\sigma = -1\,(+1)$ stands for focusing (defocusing) media.

For example, photorefractive crystals admit relatively high nonlinearity of saturable character already at intensities typical for continuous wave helium-neon lasers, and what is important the saturation level could be controlled by biasing background illumination [30-34]. Notice that in photorefractive crystals the sign of nonlinearity depends on polarity of applied static electric field. For typical photorefractive SBN crystal (electro-optic coefficient $r = 1.8\times 10^{-10}$ m/V, linear refractive index $n_0 = 2.33$) biased with dc static electric field $E_0 \sim 10^5$ V/m, for laser beams with width 10 $\mu$m at wavelength $\lambda = 0.63$ $\mu$m, propagation distance $\xi = 1$ corresponds to 2.3 mm of actual crystal length, while dimensionless amplitude $q \sim 1$ corresponds to real peak intensities about 50 mW/cm$^2$ and the typical value of saturation parameter can be estimated as $S \sim 0.2$.

We search for the simplest trivial-phase stationary periodic solutions of Eq. (1) of the form $q(\eta,\xi) = w(\eta)\exp(ib\xi)$, where $w(\eta) = w(\eta+T)$ is the real periodic function, and $b$ is the propagation constant. Upon substitution of the field in such form into Eq. (1) one gets

$$\frac{1}{2}\frac{d^2 w}{d\eta^2} - \frac{\sigma w^3}{1+Sw^2} - bw = 0. \tag{2}$$

The propagation constant $b$ is directly related to the energy flow

$$U = \int_{-T/2}^{T/2} w^2(\eta) d\eta \,, \quad (3)$$

inside each transverse wave period. The scaling analysis of Eq. (1) leads to the conclusion that if $q(\eta, \xi, S)$ is a solution of Eq.(1), then $\chi q(\chi\eta, \chi^2\xi, \chi^{-2}S)$ is also the a solution, where $\chi > 0$ is an arbitrary scaling factor. Since one can use this transformation to get various periodic wave families from the known ones, below we choose the transverse scale so that wave period $T = 2\pi$ and vary the propagation constant. The integral width defined by

$$W = 2\left( \int_{-T/4}^{T/4} w^2(\eta)\eta^2 d\eta \right)^{1/2} \left( \int_{-T/4}^{T/4} w^2(\eta) d\eta \right)^{-1/2} \quad (4)$$

is an important characteristic of the solution that defines localization of the energy inside each wave period.

The basic properties of periodic solutions of Eq. (2) are summarized in Figs. 1(a),(b) for dark sn-type wave ($\sigma = +1$); Figs. 2(a),(b) for bright cn-type wave ($\sigma = -1$); and in Figs. 3(a),(b) for bright dn-type waves ($\sigma = -1$). Sn- and cn-type waves periodically change their sign (Figs. 1(b), 2(b)), whereas dn-wave is always positive and contains a constant pedestal (Fig. 3(b)). Though exact solutions of Eq. (2) are described by Jacoby elliptic functions of sn, cn, and dn types only in the limiting case $S \to 0$ [27-29], here we by analogy keep these notations since saturation of nonlinear response does not change the topology of the solutions [35].

The dispersion diagrams (energy flow versus the propagation constant) of sign-altering sn- and cn-waves are quite similar (see Figs. 1(a), 2(a)). In the low-energy limit $U \to 0$ as well as in the high-energy limit $U \to \infty$ both sn- and cn-waves transform into harmonic patterns. The corresponding low- and high-energy cut-offs are given by $b_{w \to 0} = -1/2$ and $b_{w \to \infty} = -1/2 - \sigma / S$. The domain of existence of dn-type wave is narrowed with growth of saturation parameter (Fig. 3(a)), and near cut-offs dn-wave is close to plane one. For strong localization (especially for small values of $S$) at intermediate energies cn-waves transform into arrays of out-of-phase bright solitons, dn-wave into arrays of in-phase bright solitons, and sn-wave into arrays of out-of-phase kinks, or dark solitons. The integral width reaches its maximum value for sn-waves and its minimum value for cn-waves at intermediate energy levels that corresponds to the narrowest dark holes and bright peaks for sn- and cn-waves, respectively.

To perform the linear stability analysis of periodic waves in the saturable medium we use the new mathematical formalism, initially developed for periodic waves in cubic nonlinear media [27-29]. We search for perturbed solution of Eq. (1) in the form:

$$q(\eta,\xi) = [w(\eta) + U(\eta,\xi) + iV(\eta,\xi)]\exp(ib\xi), \tag{5}$$

with $w(\eta)$ being the stationary solution of Eq. (2), and $U, V$ real and imaginary parts of small perturbation, respectively. We search for exponentially growing perturbations $U(\eta,\xi) = \mathrm{Re}[\,u(\eta,\delta)\exp(\delta\xi)\,]$, $V(\eta,\xi) = \mathrm{Re}[\,v(\eta,\delta)\exp(\delta\xi)\,]$, where $\delta$ is the complex growth rate. Substitution of (5) into (1) and linearization yields the matrix equation for the perturbation eigenvector $\Phi(\eta) = \{u, v, du/d\eta, dv/d\eta\}^{\mathrm{T}}$:

$$\frac{d\Phi}{d\eta} = \mathcal{B}\Phi, \quad \mathcal{B} = \begin{pmatrix} \mathcal{O} & \mathcal{E} \\ \mathcal{N} & \mathcal{O} \end{pmatrix},$$

$$\mathcal{N} = \begin{pmatrix} 2b + 2\sigma\dfrac{3w^2 + Sw^4}{(1+Sw^2)^2} & 2\delta \\ -2\delta & 2b + 2\sigma\dfrac{w^2 + Sw^4}{(1+Sw^2)^2} \end{pmatrix}, \tag{6}$$

where $\mathcal{O}$ and $\mathcal{E}$ are zero and unity $2\times 2$ matrices, respectively. The general solution of Eqs. (6) can be written as $\Phi(\eta) = \mathcal{J}(\eta,\eta')\Phi(\eta')$, where $\mathcal{J}(\eta,\eta')$ is the Cauchy matrix, that is the solution of the initial value problem $\partial\mathcal{J}(\eta,\eta')/\partial\eta = \mathcal{B}(\eta)\mathcal{J}(\eta,\eta')$, $\mathcal{J}(\eta',\eta') = \mathcal{E}$.

The Cauchy matrix defines the matrix of translation of the perturbation eigenvector $\Phi$ on one wave period, as $\mathcal{P}(\eta) = \mathcal{J}(\eta + T, \eta)$. It was rigorously proven in Ref. [27-29], that the perturbation eigenvector $\Phi_k(\eta)$ is finite along the transverse $\eta$-axis when the corresponding eigenvalue of the matrix of translation fulfils the condition $|\lambda_k| = 1$ $(k = 1,\ldots,4)$. Namely this condition defines the algorithm of search for the areas of existence of “allowed” perturbations. The eigenvalues $\lambda_k$ are defined by the characteristic polynomial $D(\lambda) = \det(\mathcal{P} - \lambda\mathcal{E}) = \sum_{k=0}^{4} p_k \lambda^{4-k} = 0$. The coefficients of the polynomial are given by the traces of translation matrix $T_k = \mathrm{Tr}[\mathcal{P}^k(\eta)]$. One finds that $p_0 = p_4 = 1$, $p_1 = p_3 = -T_1$, $p_2 = (T_1^2 - T_2)/2$. Two of the four eigenvalues $\lambda_k$ can be excluded because $\lambda_k = 1/\lambda_{k+2}$ $(k = 1,2)$, and corresponding eigenvectors fulfill the symmetry relations $\Phi_k(\eta) = \Phi_{k+2}(-\eta)$. Notice that the method used here is advantageous in comparison with standard eigenvalue solvers since only $4\times 4$ matrices are required for the construction of the areas of existence of finite perturbations.

We have considered perturbations of sn-, cn- and dn-waves with general complex growth rates and explored a broad interval of saturation parameters $0 \le S \le 10$. The whole complex $\delta$-plane was covered with the fine grid (typically the step in the modulus of $\delta$ was 0.001 and the step in the phase was $\pi/1000$) and thoroughly scanned. For periodic wave patterns the areas of existence of finite perturbations have a band structure. If wave is stable condition $|\lambda_{1,2}| = 1$ is fulfilled only for imaginary $\delta$.

In contrast to the case of defocusing cubic medium, where sn-waves are stable in the entire domain of their existence, we have found that in saturable media sn-waves become weakly unstable when their energy flow exceeds a certain critical value (i.e., when $b \ge b_{\mathrm{cr}}$). This instability corresponds to complex growth rates and, hence, is of the oscillatory type. The maximum real part of the complex growth rate versus propagation constant is shown in Fig. 1(c) for various values of the saturation parameter. Notice that the oscillatory instability of the sn-waves occurs in the narrow band of propagation constants near the high-energy cut-off. Areas of existence of stable and unstable sn-waves are shown in Fig. 1(d).

Cn-waves in cubic media suffer from oscillatory instabilities in their entire domain of existence. In contrast, the central result of the present section is that cn-waves in focusing saturable media become linearly stable when their energy flow exceeds a certain critical level (Figs. 2(c) and 2(d)). This is consistent with the strong stabilizing action of saturation of nonlinear response on propagation of self-sustained nonlinear waves, dramatically illustrated by the stabilization of bell-shaped bright soliton in bulk media. The oscillatory-type instability of cn-waves in saturable media occurs for the band of propagation constants near the low-energy cut-off, i.e. for $-1/2 \le b \le b_{\mathrm{cr}}$. The maximal real part of complex growth rate versus propagation constant is shown in Fig. 2(c). Notice that real parts of growth rates quickly decrease with growth of saturation parameter. This means that even periodic waves from the unstable region can potentially be observed at high saturation levels because their typical decay length would exceed any experimentally feasible crystal length. Notice that the width of the stability band for cn-waves increases with decrease of saturation parameter. In the limit $S \to 0$ results are in full agreement with those for unstable cn-waves in cubic media.

Dn-type waves were found to suffer from exponentially growing instabilities in the entire domain of their existence – the result that is typical for the waves with pedestal. Areas for finite perturbations with real growth rates are shown in Fig. 3(c). It is interesting that the structural instability of dn-wave typically manifests itself in fusion of neighboring peaks (Fig. 3(d)), just as in the case of interaction of two in-phase solitons [36].

To confirm the results of the linear stability analysis and to elucidate its actual impact in the long-term evolution of the periodic waves, we have also integrated Eq. (1) using the beam propagation method with input conditions $q(\eta, \xi = 0) = w(\eta)[1 + \rho(\eta)]G(\eta)$, where $w(\eta)$ is the profile of the stationary wave, $\rho(\eta)$ is a Gaussian noise with the variance $\sigma^2_{\text{noise}}$, and $G(\eta)$ describes a broad Gaussian envelope imposed on the infinite periodic pattern. The width of the envelope was much larger than the wave period. The results of simulations are in full agreement with results of linear stability analysis. For example, Figs. 4(a)-4(d) illustrate the stable propagation of perturbed snoidal and cnoidal waves. The periodic waves maintain their input structure over several thousand units that exceed any feasible crystal length by several orders of magnitude. For instance in typical photorefractive crystals propagation up to 1000 units shown in Figs. 4(a)-4(d) corresponds to actual physical distance about 2 meters. In some simulations we used quite big noise variance (up to $\sigma^2_{\text{noise}} = 0.06$) to address the stability of strongly perturbed periodic patterns of sn- and cn-types, and found that they can survive even in this case.

## 3. Periodic waves in quadratic nonlinear media

In this section we consider the case of light pulse (beam) propagation in a quadratic nonlinear crystal where a fundamental frequency wave (FF) and its second harmonic (SH) interact with each other near phase-matching. The evolution of the corresponding slowly varying envelopes under conditions for non-critical type I phase matching is described by the system of coupled equations [37-40]:

$$
\begin{aligned}
i\frac{\partial q_1}{\partial \xi} &= \frac{d_1}{2}\frac{\partial^2 q_1}{\partial \eta^2} - q_1^* q_2 \exp(-i\beta\xi), \\
i\frac{\partial q_2}{\partial \xi} &= \frac{d_2}{2}\frac{\partial^2 q_2}{\partial \eta^2} - q_1^2 \exp(i\beta\xi),
\end{aligned} \tag{7}
$$

where the transverse $\eta$ coordinate is scaled in terms of the characteristic pulse (beam) width, the longitudinal $\xi$ coordinate is scaled with the dispersion (diffraction) length, parameters $d_{1,2}$ stand for dispersive (diffractive) properties of the medium for FF and SH waves, respectively, and parameter $\beta$ describes phase mismatch. We assume that there is no walk-off between FF and SH waves, as in non-critical phase-matching.

By analogy with the case of saturable medium we search for the stationary solutions of Eqs. (7) in the form $q_{1,2}(\xi,\eta) = w_{1,2}(\eta)\exp(ib_{1,2}\xi)$, where $w_{1,2}(\eta)$ are real functions, and $b_{1,2}$ are real propagation constants which should satisfy $b_2 = \beta + 2b_1$. The resulting system of equations for $w_{1,2}(\eta)$ can be written as:

$$
\begin{aligned}
\frac{d_1}{2}\frac{d^2 w_1}{d\eta^2} + b_1 w_1 - w_1 w_2 &= 0, \\
\frac{d_2}{2}\frac{d^2 w_2}{d\eta^2} + (\beta + 2b_1) w_2 - w_1^2 &= 0,
\end{aligned} \tag{8}
$$

which must be solved with periodic boundary conditions $w_{1,2}(\eta) = w_{1,2}(\eta + T)$. Notice that particular examples of periodic wave solutions of Eqs. (8) can be obtained analytically with the aid of Hamiltonian formalisms, direct substitution, and Lie group techniques [41-43]. However, whole families of solutions should be obtained numerically [44-46]. In the large phase-mismatch limit $|\beta| \gg 1$ Eqs. (8) can be converted into single cubic Schrodinger equation for the FF wave having three different types of solutions in the form of sn-, cn-, and dn-type waves. Therefore the general solutions of Eqs. (8) at moderate mismatches can be classified in accordance with their asymptotical sn-, cn-, or dn-type shape in the large mismatch limit.

Continuous families of bright and dark soliton solutions of Eqs. (8) correspond to two qualitatively different situations. For $d_1 d_2 < 0$ (this is possible for temporal solitons only) Eqs. (8) admit of continuous families of dark soliton solutions, whereas situation $d_1 d_2 > 0$ (this can be achieved for both spatial and temporal solitons) corresponds to bright soliton families.

We start from the former case (corresponding to sn-type waves) and set for simplicity $d_1 = 1$, $d_2 = -1$. Properties of sn-type waves are summarized in Fig. 5. Since one can use scaling transformations $q_{1,2}(\eta,\xi,\beta) \rightarrow \chi^2 q_{1,2}(\chi\eta,\chi^2\xi,\chi^2\beta)$ to obtain snoidal waves with different periods from a given family, we selected the transverse scale in such way that the period $T = 2\pi$. Fig. 5(a) shows energy flow $U = \int_{-T/2}^{T/2} (w_1^2 + w_2^2) d\eta$ per period versus propagation constant $b_1$. There are cut-off values for the propagation constant at different phase mismatches. Thus for $-1 \le \beta < \infty$ one has $b_1 \ge 1/2$ and both FF and SH waves disappear as $b_1 \rightarrow 1/2$, while for $\beta < -1$ propagation constant $b_1 \ge -\beta/2$ and $w_1 \rightarrow 0$, $w_2 \rightarrow -(1+\beta)/2$ in the cut-off point. Among the important characteristics of the dark periodic waves is the contrast

$$C_{1,2} = \frac{|w_{1,2}|_{\max} - |w_{1,2}|_{\min}}{|w_{1,2}|_{\max} + |w_{1,2}|_{\min}}, \tag{9}$$

which is directly related to the amplitude of the constant background in SH wave profile, and, hence, to the potential stability/instability of dark waves. The closer the contrast to one, the smaller is the amplitude of the constant background. Notice that the FF wave does not contain constant background and thus $C_1 \equiv 1$ always, thus in Fig. 5(b) we show only the contrast for the SH wave. At $\beta > 0$ the contrast $C_2$ varies only slightly, whereas for $\beta < 0$ it rapidly grows near the cut-off point and saturates to the constant limit with growth of the propagation constant (or energy flow). Notice that the higher is the phase mismatch the higher is the contrast. In a quadratic medium the integral width can be defined as:

$$W = 2\left(\int_{-T/4}^{T/4} (w_1^2 + w_2^2)\eta^2 d\eta\right)^{1/2} \left(\int_{-T/4}^{T/4} (w_1^2 + w_2^2) d\eta\right)^{-1/2}. \tag{10}$$

For sn-type waves at $\beta \ge 0$ this width slowly increases with the increase of the energy flow indicating that dark intensity holes become narrower for high-energy snoidal waves. Another important characteristic of periodic waves in quadratic nonlinear media is the energy sharing

$$S_{1,2} = \frac{1}{U} \int_{-T/2}^{T/2} w_{1,2}^2 d\eta \tag{11}$$

between the FF and the SH waves. It is shown in Fig. 5(c) as a function of $b_1$ for one particular case, at negative phase mismatch. At low energy levels and negative phase mismatches the main part of the energy is concentrated in the SH wave, whereas at high energies $S_1$ is higher than $S_2$. In the case of positive mismatch the FF wave always contains the largest part of the total energy. The typical profile of the dark snoidal periodic wave is depicted in Fig. 5(d). Notice that sn-type waves constitute the periodic analogs of the bound states of a few dark solitons studied in Refs. [47,48].

For stability analysis of periodic waves in quadratic nonlinear media we seek for perturbed solutions of Eqs. (7) in the form

$$q_{1,2}(\eta,\xi) = [w_{1,2}(\eta) + U_{1,2}(\eta,\xi) + iV_{1,2}(\eta,\xi)]\exp(ib_{1,2}\xi), \tag{12}$$

with $U_{1,2}(\eta,\xi) = \mathrm{Re}[u_{1,2}(\eta,\delta)\exp(\delta\xi)]$ and $V_{1,2}(\eta,\xi) = \mathrm{Re}[v_{1,2}(\eta,\delta)\exp(\delta\xi)]$. As in the saturable medium linearization around stationary solution yields the matrix equation for perturbation vector $\Phi(\eta) = \{u_1, u_2, v_1, v_2, du_1/d\eta, du_2/d\eta, dv_1/d\eta, dv_2/d\eta\}^{\mathrm{T}}$:

$$\frac{d\Phi}{d\eta} = \mathcal{B}\Phi, \quad \mathcal{B} = \begin{pmatrix} \mathcal{O} & \mathcal{E} \\ \mathcal{N} & \mathcal{O} \end{pmatrix},$$
$$\mathcal{N} = \begin{pmatrix} -2(b_1 - w_2)/d_1 & 2w_1/d_1 & -2\delta/d_1 & 0 \\ 4w_1/d_2 & -2b_2/d_2 & 0 & -2\delta/d_2 \\ 2\delta/d_1 & 0 & -2(b_1 + w_2)/d_1 & 2w_1/d_1 \\ 0 & 2\delta/d_2 & 4w_1/d_2 & -2b_2/d_2 \end{pmatrix}, \tag{13}$$

where $\mathcal{O}$ and $\mathcal{E}$ are zero and unity $4\times 4$ matrices, respectively. Following the procedure described in Section 2 one can construct the areas of existence of finite perturbations defined by conditions $|\lambda_k| = 1$ $(k = 1,\ldots,8)$. Eigenvalues $\lambda_k$ are roots of characteristic polynomial $\det(\mathcal{P} - \lambda\mathcal{E}) = \sum_{k=0}^{8} p_k \lambda^{8-k} = 0$. Polynomial coefficients are given by $p_0 = 1$, $p_1 = -T_1$, $p_2 = (T_1^2 - T_2)/2$, $p_3 = -T_1^3/6 + T_1T_2/2 - T_3/3$, $p_4 = T_1^4/24 - T_1^2T_2/4 + T_1T_3/3 + T_1^2/8 - T_4/4$, $p_5 = p_3$, $p_6 = p_2$, $p_7 = p_1$, $p_8 = p_0$. Symmetry relations reads now as $\lambda_k = 1/\lambda_{k+4}$ and $\Phi_k(\eta) = \Phi_{k+4}(-\eta)$ $(k = 1,...,4)$.

Stability analysis revealed that at $\beta \le -1$ sn-type waves are exponentially unstable in the whole domain of their existence, whereas for $\beta \ge -1$ exponential instabilities were found inside certain band of propagation constants (Fig. 6(a)). The areas of existence of finite perturbations at negative and positive phase mismatches are shown in Figs. 6(b) and 6(c). Notice that at $\beta \le -1$ the instability growth rates are quite high and thus will lead to a fast decay of the corresponding wave in the whole range of its existence. The important result revealed here is that at $\beta \ge -1$ the situation drastically changes: While exponential instabilities are still possible in a narrow band of propagation constants (Fig. 6(a)), the maximal value of the corresponding growth rate inside this band is very small compared with typical growth rates encountered when $\beta \le -1$. Moreover, when the mismatch $\beta$ increases, the maximum value of the growth rate inside the instability band quickly drops off. This is consistent with the suppression of exponential instabilities of sn-type waves in the large positive phase mismatch limit, i.e., in effective Kerr medium. Therefore, on physical grounds the important implication is thus near phase-matching, which is the most interesting case experimentally, such small growth rate of the exponential instabilities can only manifest itself after a long propagation distance, typically far larger than any feasible quadratic crystal length.

The "oscillatory" instabilities of sn-waves were found to exist for all energy levels and material parameters. At negative phase mismatches the oscillatory instability is strong and the real parts of the complex growth rates reach almost the same magnitude as those associated with exponential instabilities. However at positive and zero phase mismatches the maximum positive value of the complex growth rates $\mathrm{Re}(\delta)_{\max}$ associated with oscillatory instabilities was found to be quite small and to grow monotonically with the energy flow $U$. Notice that the band of energy flows where $\mathrm{Re}(\delta)_{\max} \le 0.05$ rapidly increases from $0 \le U \lesssim 62.2$ at $\beta = 3$ to $0 \le U \lesssim 314$ at $\beta = 10$ (Fig. 6(d)). Since decay length of the waves with such low increments is huge the important conclusion is that it should be possible to observe sn-type waves experimentally. In the limit $\beta \to \infty$ the "metastability" region broadens and both oscillatory and exponential instabilities are suppressed.

We have found that dominant frequencies in the spectrum of perturbation with highest $\mathrm{Re}(\delta)$ could be estimated as $\Omega \approx \pm[2\,\mathrm{Im}(\delta)]^{1/2}$ and at moderate and high energy flows are much higher than frequencies of the own harmonics of snoidal waves. This means that in the case of well-localized high-energy waves the most "harmful" high-frequency perturbations lay far from the frequency band of snoidal waves, and could be potentially removed by spectral filtering. For example for a wave with $b_1 = 7$ at $\beta = 3$ the frequency band of the snoidal wave is given by $-8 \lesssim \Omega \lesssim 8$ while the dominant frequencies in perturbation spectrum is $\Omega \approx \pm 24$ (see Fig. 6(e) which shows the profile of the perturbation for this wave). A typical scenario of the instability development for the snoidal-type waves is shown in Fig. 6(f). One can clearly see appearance of high frequency modulations of the otherwise smooth profile, however the depth of this modulation is small and the wave conserves structural stability.

Next we concentrate on the properties of dn- and cn-type waves in quadratic nonlinear media. Such waves feature arrays of bright solitons and exist at $d_1 d_2 > 0$. Therefore further we set in Eqs. (8) $d_1 = -1$, $d_2 = -1/2$, that corresponds for instance to spatial soliton case. Examples of dispersion diagrams and profiles of dn- and cn-type waves are shown in Figs. 7(a),(b) and 8(a),(b), respectively. Waves of cn- and dn-types exist for propagation constant values above a cut-off. For cn-waves the cut-off is given by $b_1 = -\beta/2$ at $-\infty < \beta \le 1$, and by $b_1 = -1/2$ at $\beta > 1$. The cut-off for dn-waves is always positive; it monotonically decreases with increase of the phase mismatch, and approaches $b_1 = -\beta/2$ as $\beta \to -\infty$ [44]. In the high-energy, or high-localization limit, cn-type waves transform into arrays of out-of-phase bright solitons, while dn-type waves transform into arrays of in-phase bright solitons. In the low energy limit, dn-type waves transform into plane waves, whereas cn-type waves transform either into plane or small amplitude harmonic waves depending on the phase mismatch sign.

Stability analysis revealed that dn-type waves are linearly unstable in the entire domain of their existence. The areas of existence of finite perturbations corresponding to real $\delta$ at $\beta = 0$ are depicted in Fig. 7(c). The exponentially growing instability is most pronounced at the low-energy limit, when $b_1$ approaches cut-off and the dn-wave transforms into a plane wave, and it is asymptotically suppressed in the high-energy limit, when the dn-wave transforms into an array of in-phase bright solitons. The typical scenario of destabilization of a perturbed dn-wave is shown in Fig. 7(d). The instability of dn-wave always manifests itself as fusion of neighboring peaks. Similar results were obtained for all values of the propagation constant at different phase-mismatches inside the interval investigated, namely $-20 \le \beta \le 20$.

Next we consider stability of cn-type waves. On intuitive grounds, such waves are expected to be more robust than the dn-waves because of the alternating phase of the FF, hence neighboring peaks might tend to repel each other and thus a stable balance might be possible. Still, such is not the case, e.g., in self-focusing cubic Kerr-type nonlinear media where the cn-waves are unstable, too. The central result of this section is that such stable balance does occur with multicolor cn-waves, as shown in Fig. 8(c)-(f). The dynamically saturable nature of quadratic nonlinearities might be important to such stabilization. The important result, shown in Figs. 8(c) and 8(d), is that the areas of existence of growing perturbations ($\mathrm{Re}(\delta) > 0$) vanish when the propagation constant exceeds a certain critical value $b_{\mathrm{cr}}$. Taking into account that dispersion dependencies are monotonic (Fig. 8(a)), it means that above the corresponding energy threshold, the families of multicolor cn-waves become completely free of linear instabilities with exponential growth. Fig. 8(f) indicates that the threshold energy for stabilization decreases when the phase mismatch $\beta$ goes from $-\infty$ to approximately 0.25. Thus, stable cn-waves occur almost in the whole range of its existence at small positive and negative $\beta$. For example, at $\beta = -3$ we have found stable cnoidal waves for $b_1 \geq b_{\mathrm{cr}} \approx 2.382$, while at $\beta = 0$ the region of stability begins at $b_{\mathrm{cr}} \approx 0.192$. The exponential instabilities of low-energy cn-waves associated with purely real growth rates exist at $-\infty < \beta \leq 1$.

In the region $0.25 \lesssim \beta < \infty$ we have also found oscillatory instabilities associated with complex growth rates. Such instabilities also cease to exist when the energy flow exceeds a threshold value (Fig. 8(e)). For oscillatory instabilities the critical value of the propagation constant also decreases as $\beta$ goes from $\infty$ to approximately 0.25. The key result of this section is thus summarized in Fig. 8(f), which shows the threshold value of the propagation constant for complete stabilization versus phase mismatch.

To confirm the results of linear stability analysis we performed a set of simulations of Eqs. (7) with the input conditions $q_{1,2}(\eta,0) = w_{1,2}(\eta)G(\eta)[1 + \rho_{1,2}(\eta)]$, where $\rho_{1,2}(\eta)$ is a Gaussian random function with variance $\sigma_{1,2}^2$, and $G(\eta)$ is a wide envelope modulating the infinite periodic pattern. For sn-type waves we have found that in some cases (when growth rates predicted by linear stability analysis are small enough) perturbed sn-waves with low and moderate energy flows can survive up to one thousand propagation units exceeding any feasible crystal length by several orders of magnitude (Fig. 9). The band of energy flows corresponding to such "metastable" propagation quickly increases with growth of phase mismatch, as predicted by the stability analysis. As in the case of regular perturbations in linear stability analysis in the case of random perturbations instability of high-energy sn-waves manifests itself in appearance of high frequency oscillations and lead to the behavior analogous to that depicted in Fig. 6(f).

Figs. 10(a)-(d) illustrate the outcome of numerical simulations of propagation of perturbed cn-waves. The typical decay of low-energy cn-waves is shown in Fig. 10(a). Beyond the point of spontaneous onset of the instability, the wave is rapidly destroyed. In clear contrast, Figs 10(b)-(d) illustrate the stable propagation of perturbed cnoidal waves for several values of phase mismatch and energy flow. The cnoidal waves kept their input structure for all the monitored propagation distances, in some cases over several thousand units.

## 4. Concluding remarks

Summarizing, we have shown the existence of completely stable periodic wave patterns in both saturable Kerr-type and quadratic nonlinear media. From a theoretical viewpoint, these findings motivate the investigation of the existence and properties of periodic waves in more general models, such as in media with competing quadratic-cubic, or cubic-quintic, or in optical tandems. From a conceptual point of view, the stable periodic waves may be viewed as bright and dark “soliton crystals,” composed of individual soliton-like “atoms”. Such concept might find applications, e.g., in one-dimensional versions of soliton-based image processing techniques [49,50], or in the formation of reconfigurable light-induced periodic waveguide arrays in future photonic circuits.

## Acknowledgements

Financial support from CONACyT under the grant U39681-F is gratefully acknowledged by VAV. YVK and LT acknowledge support by the Generalitat de Catalunya and by the Spanish Government under grant TIC2000-1010.

## References

1. N. N. Akhmediev and A. Ankiewicz, 1997, Solitons (London, Chapman-Hall).
2. Yu. S. Kivshar and G. Agrawal, 2003, Optical Solitons (New York, Academic Press).
3. A. M. Kamchatnov, 2000, Nonlinear periodic waves and their modulations (Kluwer Academic Publishers).
4. E. Infeld and R. Rowlands, 1990, Nonlinear waves, Solitons and Chaos (Cambridge, Cambridge University Press).
5. H. C. Yuen and B. M. Lake, 1982, Nonlinear dynamics of deep-water gravity waves (New York, Academic Press).
6. E. A. Kuznetsov, A. M. Rubenchik, and V. E. Zakharov, 1986, Phys. Rep. **142**, 103.
7. J. W. Fleischer, M. Segev, N. K. Efremidis, and D. N. Christodoulides, 2003, Nature **422**, 147.
8. S. E. Fil'chenkov, G. M. Fraiman, and A. D. Yunakovskii, 1987, Sov. J. Plasma Phys. **13**, 554.
9. V. P. Pavlenko and V. I. Petviashvili, Sov. J. Plasma Phys. **8**, 117 (1982).
10. D. U. Martin, H. C. Yuen, and P. G. Saffman, 1980, Wave Motion **2**, 215.
11. V. P. Kudashev and A. B. Mikhailovsky, 1986, Sov. Phys. JETP **63**, 972.
12. V. M. Petnikova, V. V. Shuvalov, and V. A. Vysloukh, 1999, Phys. Rev. E **60**, 1009.
13. F. T. Hioe, 1999, Phys. Rev. Lett. **82**, 1152.
14. V. Aleshkevich, Y. Kartashov, and V. Vysloukh, 2000, Opt. Commun. **185**, 305.
15. V. Aleshkevich, Y. Kartashov, and V. Vysloukh, 2001, J. Opt. Soc. Am. B **18**, 1127.
16. V. Aleshkevich, Y. Kartashov, and V. Vysloukh, 2001, Opt. Commun. **190**, 373.
17. Y. V. Kartashov, V. A. Vysloukh, E. Marti-Panameno, D. Artigas, L. Torner, 2003, Phys. Rev. E **68**, 026613.
18. V. A. Aleshkevich, V. A. Vysloukh, and Y. V. Kartashov, 2001, Quantum Electron. **31**, 257.

19. Y. V. Kartashov, V. A. Vysloukh, and L. Torner, 2003, Phys. Rev. E **68**, 015603(R).
20. L. D. Carr, C. W. Clark, and W. P. Reinhardt, 2000, Phys. Rev. A **62**, 063610.
21. L. D. Carr, C. W. Clark, and W. P. Reinhardt, 2000, Phys. Rev. A **62**, 063611.
22. J. C. Bronski, L. D. Carr, B. Deconinck, J. N. Kutz, and K. Promislow, 2001, Phys. Rev. E **63**, 036612.
23. J. C. Bronski, L. D. Carr, R. Carretero-Gonzalez, B. Deconinck, J. N. Kutz, and K. Promislow, 2001, Phys. Rev. E **64**, 056615.
24. J. C. Bronski, L. D. Carr, B. Deconinck, and J. N. Kutz, 2001, Phys. Rev. Lett. **86**, 1402.
25. A. Ankiewicz, K.-I. Maruno, and N. Akhmediev, 2003, Phys. Lett. A **308**, 397.
26. K.-I. Maruno, A. Ankiewicz, and N. Akhmediev, 2003, Physica D **176**, 44.
27. Y. V. Kartashov, V. A. Aleshkevich, V. A. Vysloukh, A. A. Egorov, and A. S. Zelenina, 2003, Phys. Rev. E **67**, 036613.
28. Y. V. Kartashov, V. A. Aleshkevich, V. A. Vysloukh, A. A. Egorov, and A. S. Zelenina, 2003, J. Opt. Soc. Am. B **20**, 1273.
29. V. A. Aleshkevich, A. A. Egorov, Y. V. Kartashov, V. A. Vysloukh, and A. S. Zelenina, 2003, Phys. Rev. E **67**, 066605.
30. G. C. Duree, J. L. Shultz, G. J. Salamo, M. Segev, A. Yariv, B. Crosignani, P. Di Porto, E. J. Sharp, and R. R. Neurgaonkar, 1993, Phys. Rev. Lett. **71**, 533.
31. M. Segev, G. C. Valley, B. Crosignani, P. Di Porto, and A. Yariv, 1994, Phys. Rev. Lett. **73**, 3211.
32. G. Duree, G. Salamo, M. Segev, A. Yariv, B. Crosignani, P. Di Porto, and E. Sharp, 1994, Opt. Lett. **19**, 1195.
33. M. Shih, P. Leach, M. Segev, M. H. Garrett, G. Salamo, and G. C. Valley, 1996, Opt. Lett. **21**, 324.
34. G. I. Stegeman and M. Segev, 1999, Science, **286**, 1518.
35. Y. V. Kartashov, A. A. Egorov, A. S. Zelenina, V. A. Vysloukh, and L. Torner, 2003, Phys. Rev. E **68**, 065605(R).
36. J. P. Gordon, 1983, Opt. Lett. **8**, 596.
37. C. R. Menyuk, R. Schiek, and L. Torner, 1994, J. Opt. Soc. Am. B **11**, 2434.
38. G. I. Stegeman, D. J. Hagan, and L. Torner, 1996, Opt. Quantum Electron. **28**, 1691.

39. A. V. Buryak, P. Di Trapani, D. V. Skryabin and S. Trillo, 2002, Phys. Rep. **370**, 63.
40. L. Torner and A. Barthelemy, 2003, IEEE J. Quantum Electron. **39**, 22.
41. P. Ferro and S. Trillo, 1995, Phys. Rev. E **51**, 4994.
42. D. F. Parker, 1998, J. Opt. Soc. Am. B **15**, 1061.
43. S. Lafortune, P. Winternitz, and C. R. Menyuk, 1998, Phys. Rev. E **58**, 2518.
44. Y. V. Kartashov, A. A. Egorov, A. S. Zelenina, V. A. Vysloukh, and L. Torner, 2004, Phys. Rev. Lett. **92**, 033901.
45. Y. V. Kartashov, V. A. Vysloukh, and L. Torner, 2003, Phys. Rev. E **67**, 066612.
46. Y. V. Kartashov, A. A. Egorov, A. S. Zelenina, V. A. Vysloukh, and L. Torner, 2003, Phys. Rev. E **68**, 046609.
47. A. V. Buryak and Y. S. Kivshar, 1995, Phys. Rev. A **51**, R41.
48. A. V. Buryak and Y. S. Kivshar, 1995, Opt. Lett. **20**, 834.
49. A. Bramati, W. Chinaglia, S. Minardi, and P. Di Trapani, 2001, Opt. Lett. **26**, 1409.
50. S. Minardi, G. Arrighi, P. Di Trapani, A. Varanavicius, and A. Piskarskas, 2002, Opt. Lett. **27**, 2097.

## Figure captions

Figure 1. (a) Energy flow of sn-type wave versus propagation constant. (b) Profiles of sn-waves with various energy flows at $S = 0.1$. (c) Maximum real part of complex growth rate versus propagation constant. (d) Areas of existence of stable and unstable (shaded) sn-waves.

Figure 2. (a) Energy flow of cn-type wave versus propagation constant. (b) Profiles of cn-waves with various energy flows at $S = 0.1$. (c) Maximum real part of complex growth rate versus propagation constant. (d) Areas of existence of stable and unstable (shaded) cn-waves.

Figure 3. (a) Energy flow of dn-type wave versus propagation constant. (b) Profiles of dn-waves with various energy flows at $S = 0.1$. (c) Areas of existence of finite perturbations with real growth rates at $S = 0.8$ (shaded). Vertical lines in (c) stand for cut-offs. (d) Propagation of dn-wave with $b = 0.6$ and $S = 0.25$ in the presence of perturbation with growth rate $\delta = 0.1084$.

Figure 4. (a) Profile of stationary sn-wave with $b = -2$ and (b) its long-term propagation in the presence of white input noise superimposed on the stationary solution. (c) Profile of stationary cn-wave with $b = 3$ and (d) its long-term propagation in the presence of noise. Saturation parameter $S = 0.25$. Noise variance $\sigma_{\text{noise}}^2 = 0.01$.

Figure 5. (a) Dispersion curves for sn-wave at various phase mismatches. (b) Wave contrast versus propagation constant for various phase mismatches. (c) Energy sharing as function of propagation constant at $\beta = -3$. (d) Profile of snoidal wave with $U = 10$ at $\beta = 0$.

Figure 6. (a) Area of existence of exponential instabilities for dark snoidal waves (shaded). Areas of existence of finite perturbations with real growth rates at $\beta = -3$ (b) and $\beta = 0, 3, 10$ (c). (d) Maximum real part of complex growth rates versus energy flow for different phase mismatches.

(e) One of the components of the perturbation corresponding to growth rate $\delta = 0.0979 + 293.3921i$ and snoidal wave with $b_1 = 7$ at $\beta = 3$. (f) Propagation of the snoidal wave in the presence of the perturbation depicted in subfigure (e). Only the FF wave is shown.

Figure 7. (a) Energy flow of dn-wave versus propagation constant for various phase mismatches. (b) Dn-wave profile at $\beta = 0$ and $b_1 = 0.6$. (c) Areas of existence of finite perturbations with real growth rates at $\beta = 0$ (shaded). Cut-off on propagation constant in figure (c) is given by $b_1 \approx 0.341$. (d) Propagation of dn-wave with $b_1 = 0.6$ at $\beta = 0$ in the presence of the perturbation with $\delta = 0.12444$. Only the FF wave is shown.

Figure 8. (a) Energy flow of cn-wave versus propagation constant for various phase mismatches. (b) Cn-wave profile at $\beta = 0$ and $b_1 = 0.15$. Areas of existence of finite perturbations with real growth rates (shaded) at $\beta = -3$ (c) and $\beta = 0$ (d). (e) Maximal real part of complex growth rate versus propagation constant at $\beta = 3$. (f) Threshold propagation constant for stabilization versus phase mismatch.

Figure 9. Row (a) shows the profile of the stationary dark snoidal wave with $b_1 = 1$ at $\beta = 3$ and its long-term propagation in the presence of white input noise superimposed to the stationary solution. Row (b): the same as in row (a) but for the wave with $b_1 = 1.6$ and $\beta = 10$. Noise variance $\sigma_{1,2}^2 = 0.01$. Only the FF wave is shown.

Figure 10. (a) Propagation of the unstable cn-wave with $b_1 = 0.15$ at $\beta = 0$ in the presence of the perturbation with $\delta = 0.0954$. Long-distance evolution of stable cn-waves with $b_1 = 1.8$ at $\beta = -2$ (b), $b_1 = 0.5$ at $\beta = 0$ (c), $b_1 = 0.5$ at $\beta = 1$ (d) in the presence of white noise with variance $\sigma_{1,2}^2 = 0.01$. Only the FF wave is shown.

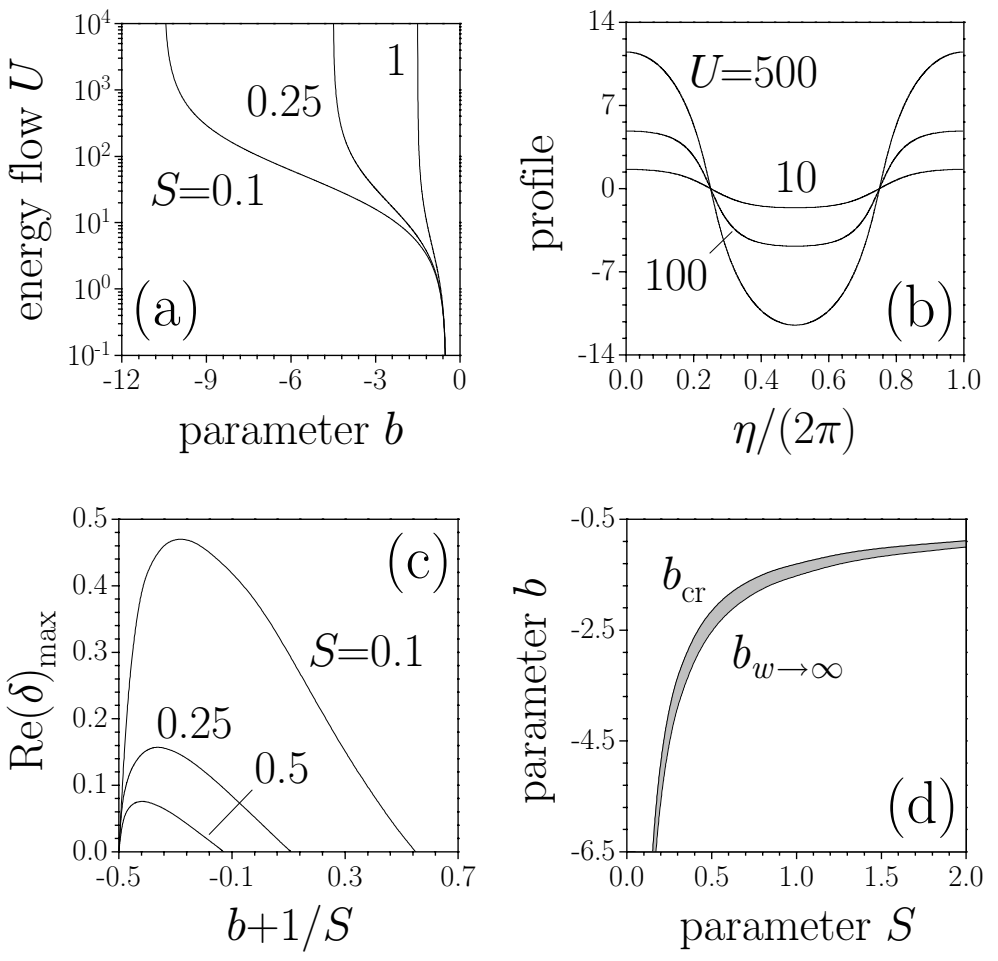

Figure 1. (a) Energy flow of sn-type wave versus propagation constant. (b) Profiles of sn-waves with various energy flows at $S = 0.1$. (c) Maximum real part of complex growth rate versus propagation constant. (d) Areas of existence of stable and unstable (shaded) sn-waves.

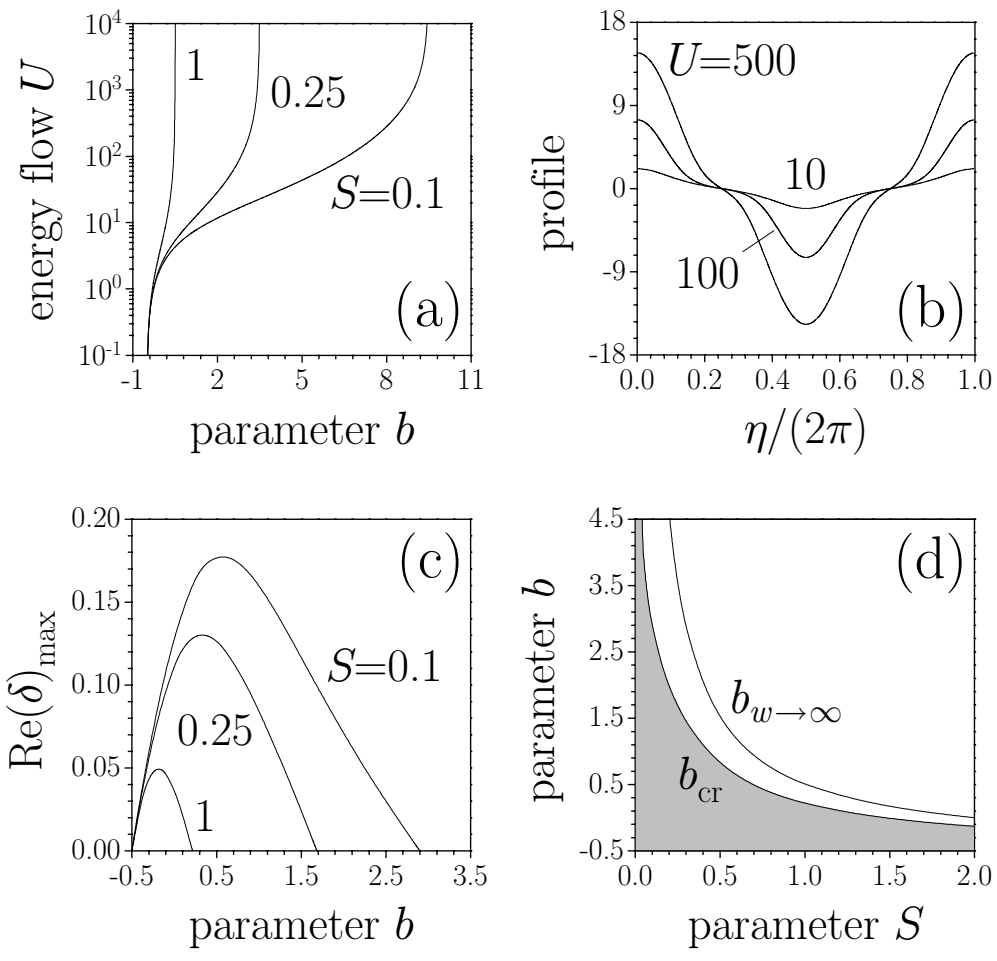

Figure 2. (a) Energy flow of cn-type wave versus propagation constant. (b) Profiles of cn-waves with various energy flows at $S = 0.1$. (c) Maximum real part of complex growth rate versus propagation constant. (d) Areas of existence of stable and unstable (shaded) cn-waves.

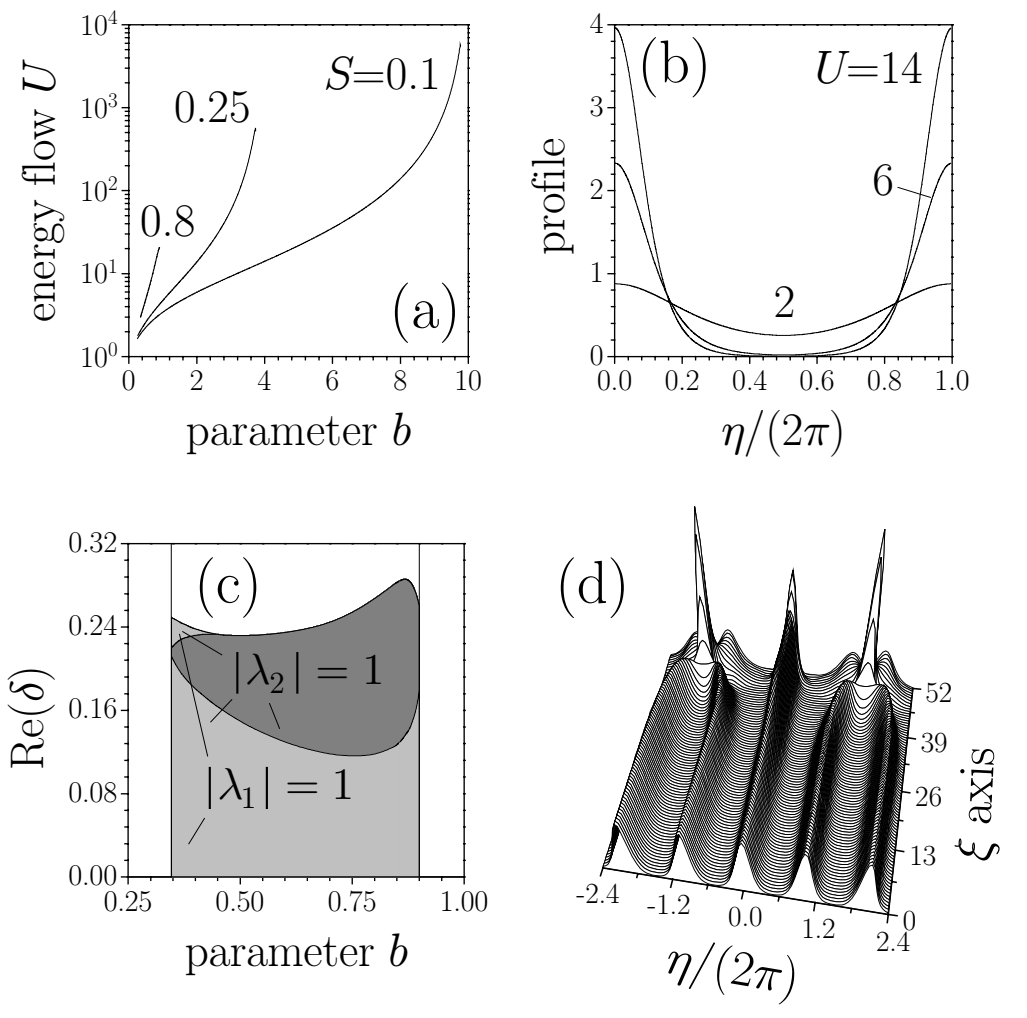

Figure 3. (a) Energy flow of dn-type wave versus propagation constant. (b) Profiles of dn-waves with various energy flows at $S = 0.1$. (c) Areas of existence of finite perturbations with real growth rates at $S = 0.8$ (shaded). Vertical lines in (c) stand for cut-offs. (d) Propagation of dn-wave with $b = 0.6$ and $S = 0.25$ in the presence of perturbation with growth rate $\delta = 0.1084$.

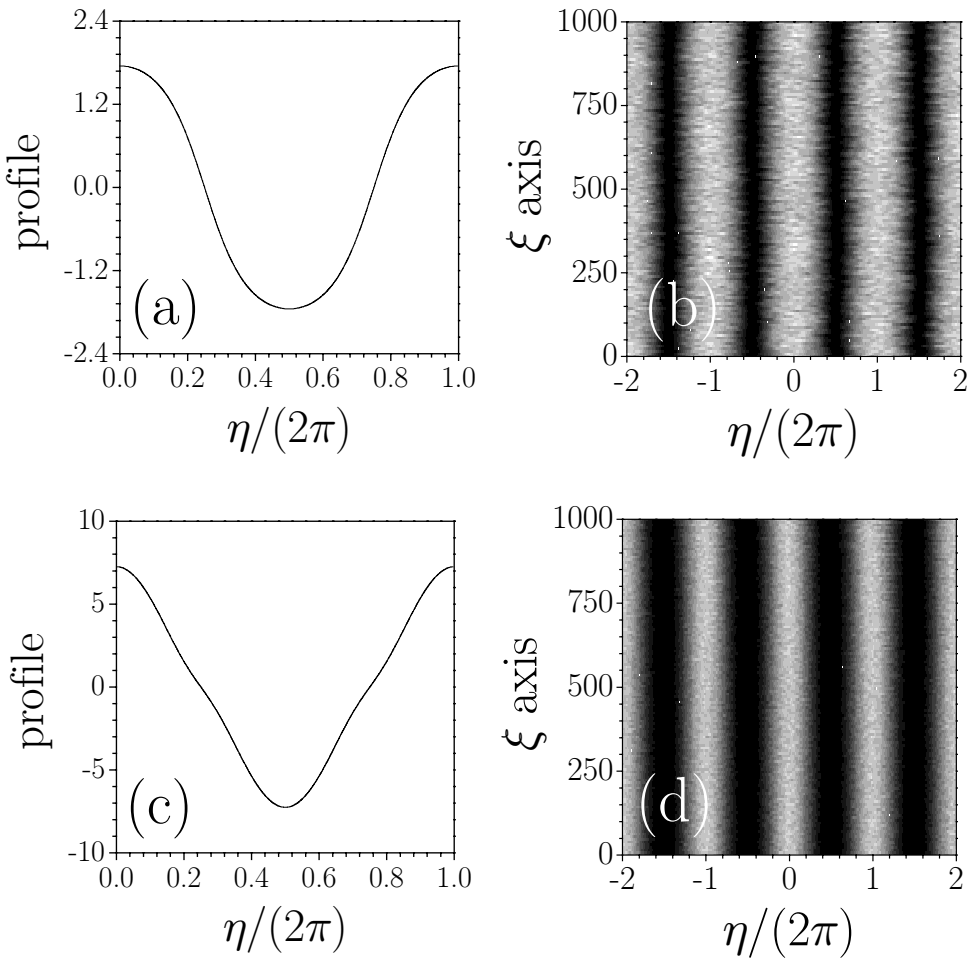

Figure 4. (a) Profile of stationary sn-wave with $b = -2$ and (b) its long-term propagation in the presence of white input noise superimposed on the stationary solution. (c) Profile of stationary cn-wave with $b = 3$ and (d) its long-term propagation in the presence of noise. Saturation parameter $S = 0.25$. Noise variance $\sigma^2_{\mathrm{noise}} = 0.01$.

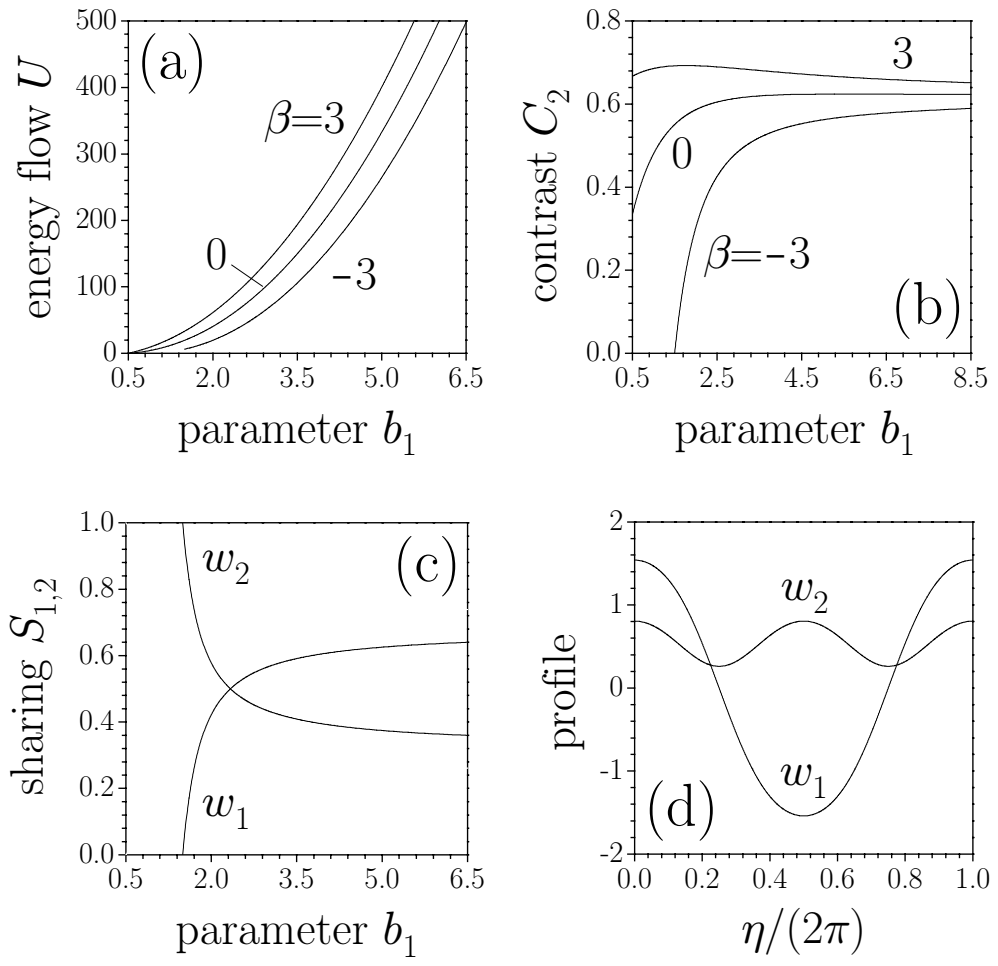

Figure 5. (a) Dispersion curves for sn-wave at various phase mismatches. (b) Wave contrast versus propagation constant for various phase mismatches. (c) Energy sharing as function of propagation constant at $\beta = -3$. (d) Profile of snoidal wave with $U = 10$ at $\beta = 0$.

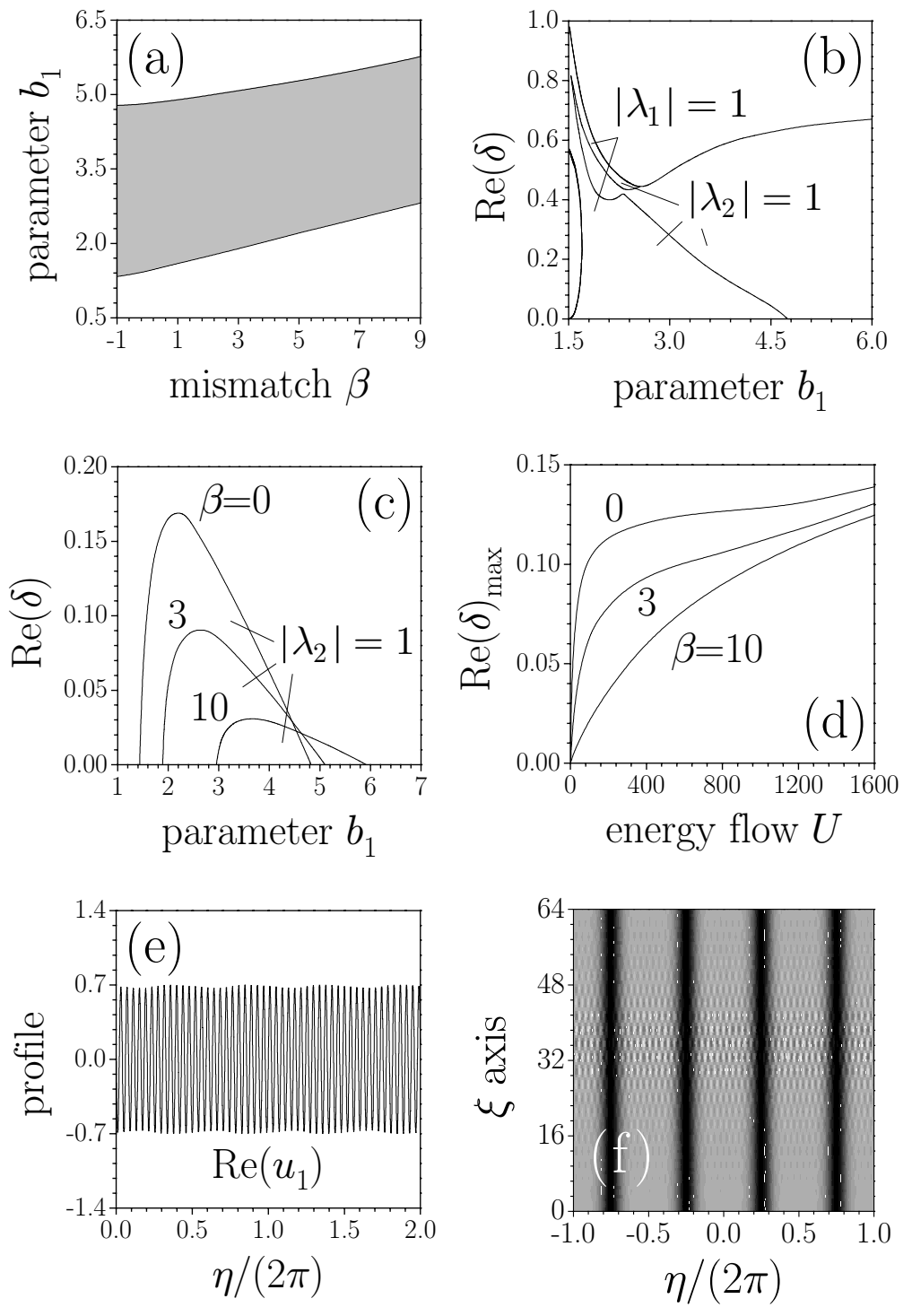

Figure 6. (a) Area of existence of exponential instabilities for dark snoidal waves (shaded). Areas of existence of finite perturbations with real growth rates at $\beta = -3$ (b) and $\beta = 0, 3, 10$ (c). (d) Maximum real part of complex growth rates versus energy flow for different phase mismatches. (e) One of the components of the perturbation corresponding to growth rate $\delta = 0.0979 + 293.3921i$ and snoidal wave with $b_1 = 7$ at $\beta = 3$. (f) Propagation of the snoidal wave in the presence of the perturbation depicted in subfigure (e). Only the FF wave is shown.

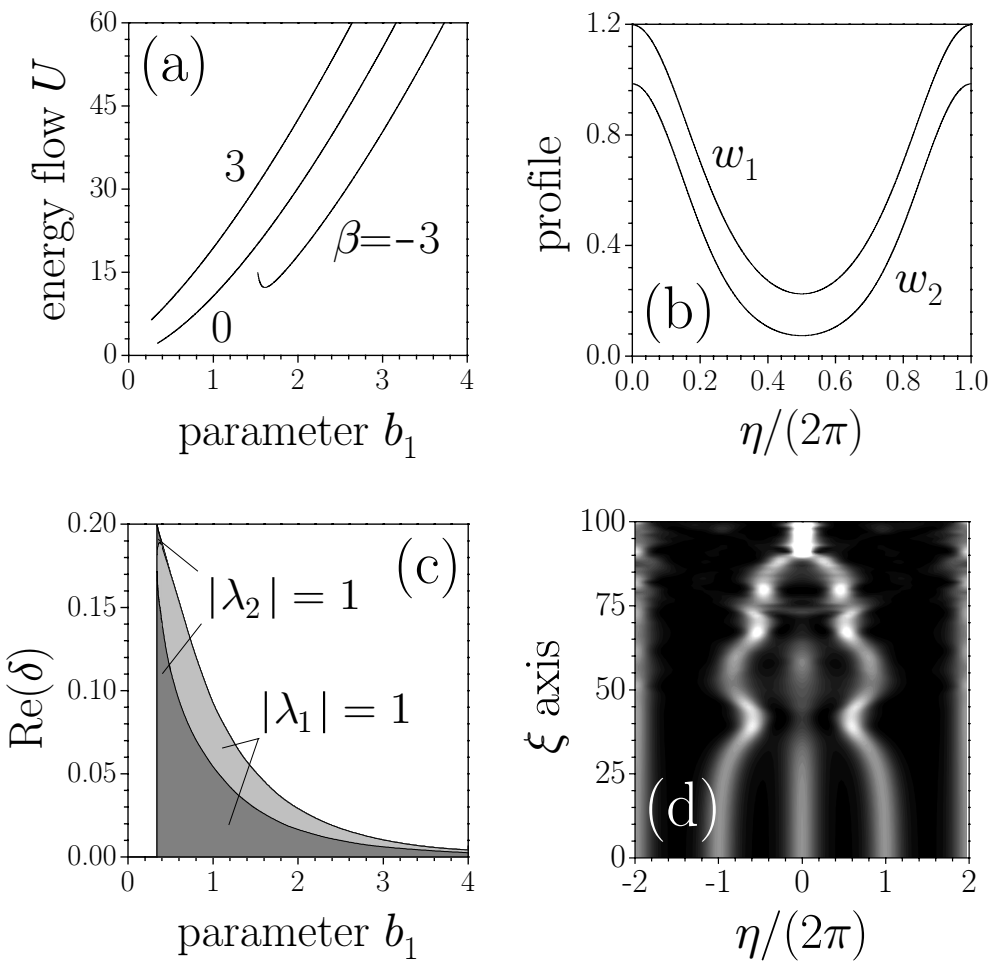

Figure 7. (a) Energy flow of dn-wave versus propagation constant for various phase mismatches. (b) Dn-wave profile at $\beta = 0$ and $b_1 = 0.6$. (c) Areas of existence of finite perturbations with real growth rates at $\beta = 0$ (shaded). Cut-off on propagation constant in figure (c) is given by $b_1 \approx 0.341$. (d) Propagation of dn-wave with $b_1 = 0.6$ at $\beta = 0$ in the presence of the perturbation with $\delta = 0.12444$. Only the FF wave is shown.

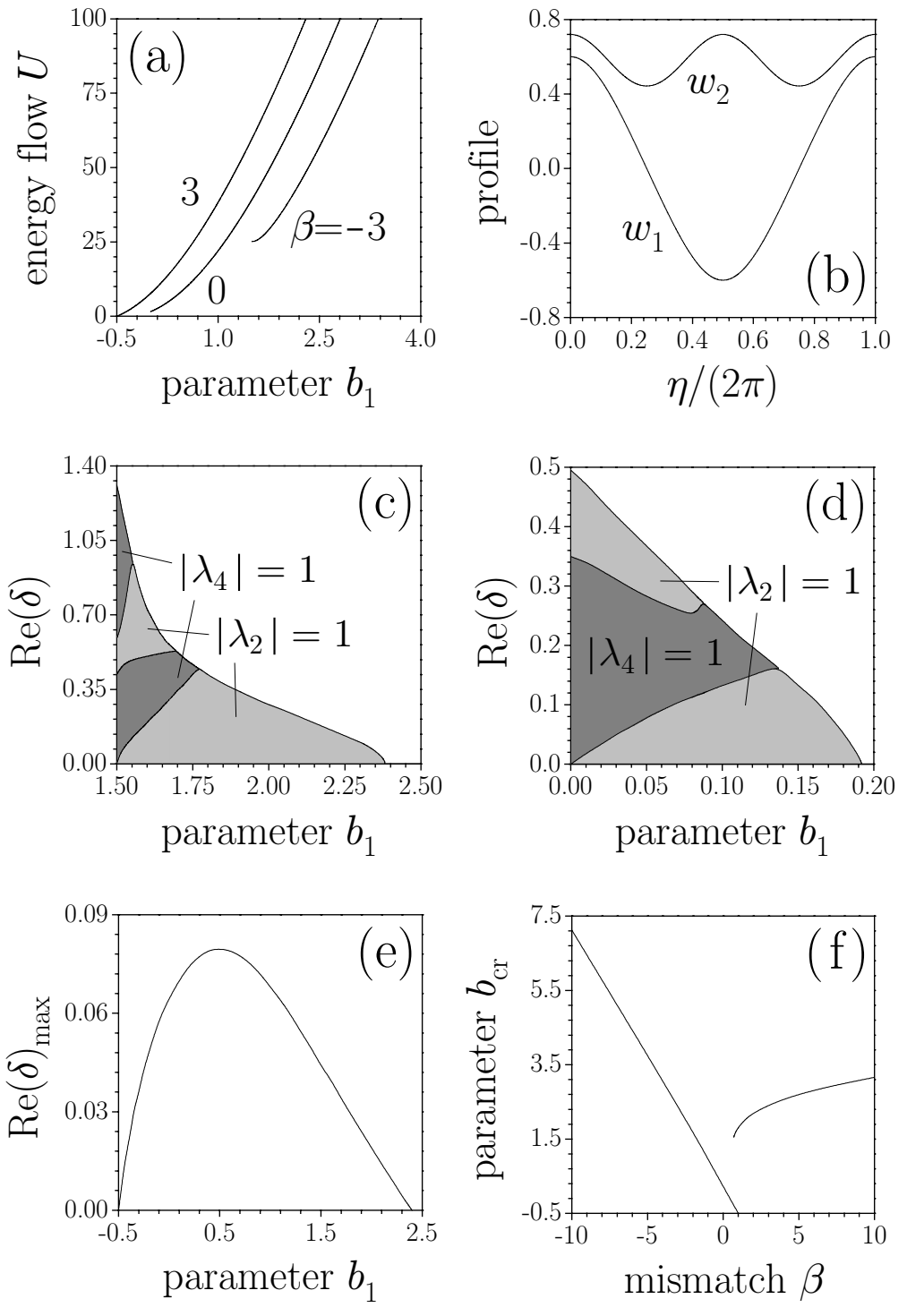

Figure 8. (a) Energy flow of cn-wave versus propagation constant for various phase mismatches. (b) Cn-wave profile at $\beta = 0$ and $b_1 = 0.15$. Areas of existence of finite perturbations with real growth rates (shaded) at $\beta = -3$ (c) and $\beta = 0$ (d). (e) Maximal real part of complex growth rate versus propagation constant at $\beta = 3$. (f) Threshold propagation constant for stabilization versus phase mismatch.

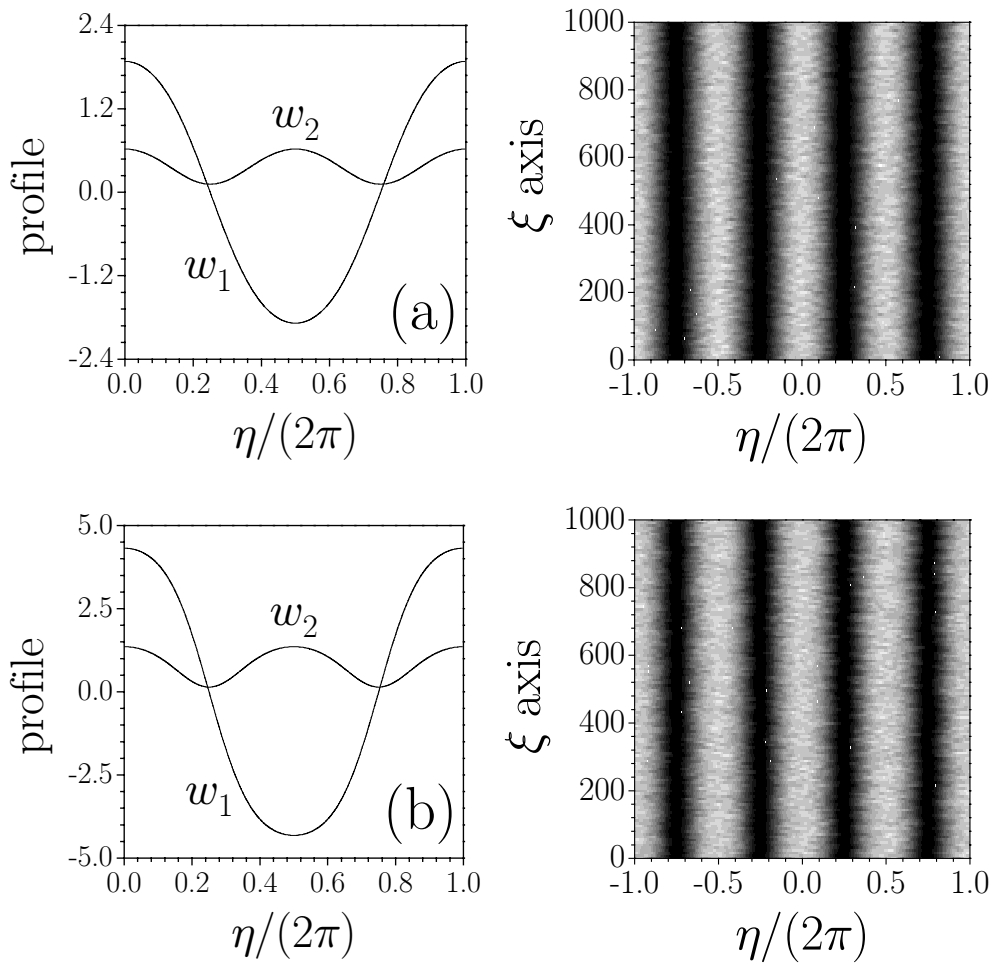

Figure 9. Row (a) shows the profile of the stationary dark snoidal wave with $b_1 = 1$ at $\beta = 3$ and its long-term propagation in the presence of white input noise superimposed to the stationary solution. Row (b): the same as in row (a) but for the wave with $b_1 = 1.6$ and $\beta = 10$. Noise variance $\sigma_{1,2}^2 = 0.01$. Only the FF wave is shown.

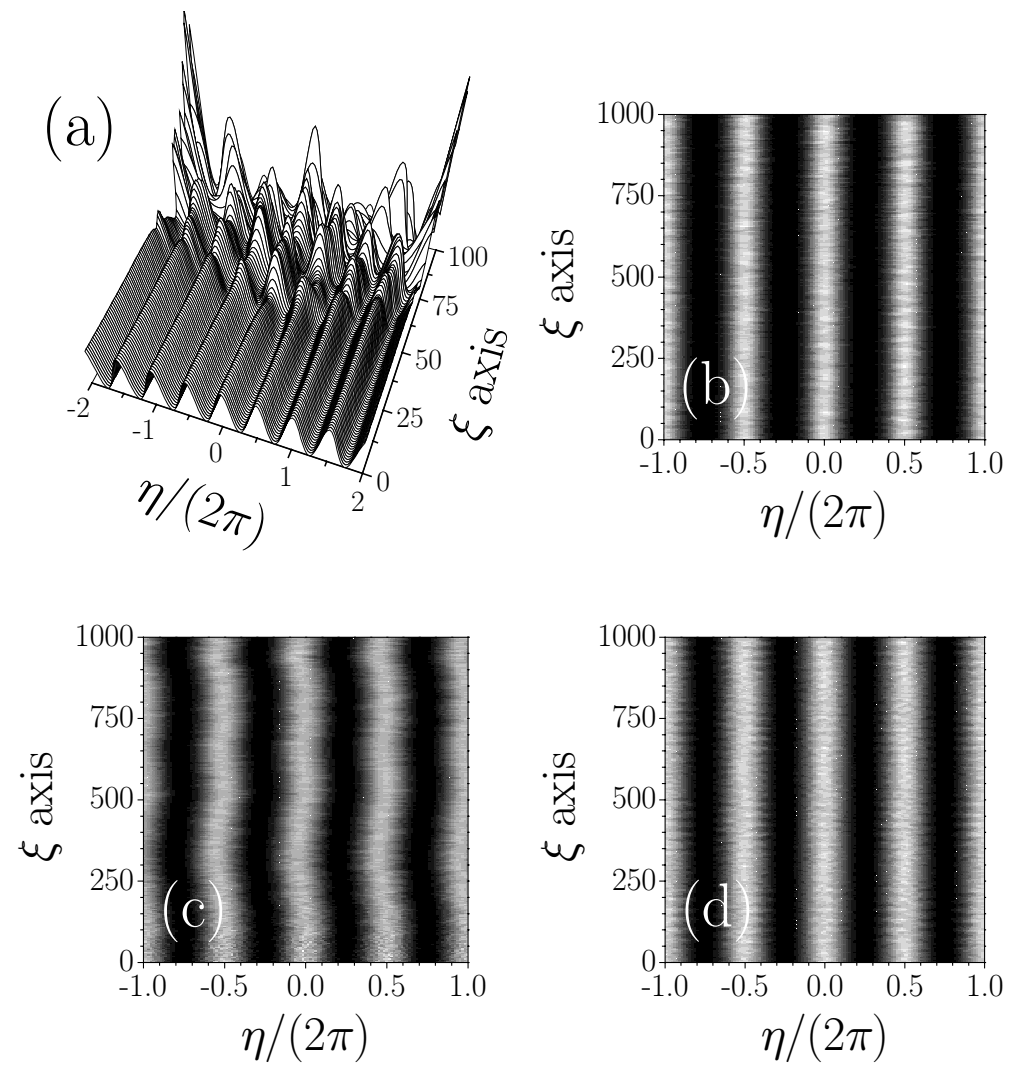

Figure 10. (a) Propagation of the unstable cn-wave with $b_1 = 0.15$ at $\beta = 0$ in the presence of the perturbation with $\delta = 0.0954$. Long-distance evolution of stable cn-waves with $b_1 = 1.8$ at $\beta = -2$ (b), $b_1 = 0.5$ at $\beta = 0$ (c), $b_1 = 0.5$ at $\beta = 1$ (d) in the presence of white noise with variance $\sigma_{1,2}^2 = 0.01$. Only the FF wave is shown.